\documentclass[aps,twocolumn,pra]{revtex4-1}
\usepackage{newlfont}
\usepackage{color}           
\usepackage{soul}            
\usepackage{amssymb}
\usepackage{amsfonts}
\usepackage{amsmath}
\usepackage{wasysym}
\usepackage{graphicx}
\usepackage{epsfig}
\usepackage{amsthm}
\usepackage{bm}
\usepackage[colorlinks=true, citecolor=blue, urlcolor=blue ]{hyperref}

\begin{document} 
 
 \title{Role of quantum correlation in metrology beyond standard quantum limit} 
\author{Manabendra N. Bera}
\affiliation{Harish-Chandra Research Institute, Chhatnag Road, Jhunsi, Allahabad 211 019, India}

\begin{abstract} 
Quantum metrology is studied in the presence of quantum correlation. The quantum correlation measure based on quantum Fisher information enables us to gain a deeper insight on how quantum correlations are instrumental in setting metrological precision.  
Our analysis shows that not only the entanglement but also the quantum correlation plays important roles to enhance precision in quantum metrology. Even in the absence of entanglement, quantum correlation can be exploited as the resource to beat standard quantum limit and reach Heisenberg limit in metrology.
Clearly unraveling the role of quantum correlations, the tighter bounds on the metrological precision are derived. 

\end{abstract}

\maketitle

\section{Introduction}  
Metrology has fundamental implications in science and technology. It is concerned with the largest possible precision achievable in various parameter estimation tasks and frame measurement schemes to achieve that precision. In metrology quantum Fisher information (QFI) plays  central role \cite{Paris09, Monras06,  Giovannetti06} and its inverse provides the lower bound on the error in statistical estimation of an unknown parameter \cite{Braunstein94, Wootters81, Petz96}. 
%
%
Hence, the ways to increase QFI become an intriguing question in quantum metrology. 
In particular, if the quantum entanglement present in a system can be used as a resource. A substantial amount of effort has been put forward in this 
context \cite{ Giovannetti06} and it has been proven that entanglement does play a positive role to enhance precision in 
metrology. Even entanglement can be exploited as the quantum resource to go beyond standard quantum limit (SQL) and attain Heisenberg limit (HL) \cite{Giovannetti04}. This fact gives rise to an important question, whether such increase in QFI can be used as the signature of quantum entanglement. 
Further, if it would be possible to relate and quantify quantum entanglement in terms of QFI. Recently 
several studies have been carried out in this direction \cite{Smerzi09}. A quantum correlation measure has been introduced, recently, based on quantum Fisher information and the measure has been shown to set 
the minimum precision achievable in ``black-box'' quantum metrology \cite{Girolami14}. However, the understanding on the role of quantum correlation in quantum metrology, in particular to attain larger precision, is far from complete. 

In this paper, 
we address two intriguing questions, relating general (pure and mixed) quantum states. 
First, if the quantum correlation other than entanglement can be the resource to enhance the precision in metrology by increasing QFI. And, if so, how do quantum 
correlations play role in it. Second, if it is possible to provide tighter bounds on the metrological precision in the presence 
of quantum correlations. 

We give affirmative answers to above questions. 
We observe, for given quantum correlated state, that not only entanglement but also quantum correlation is a resource in quantum metrology and we uncover the role behind. We demonstrate that, even in the absence of quantum entanglement, quantum correlation is capable of playing constructive role to beat SQL and attain HL.
We also provide, solely quantum state dependent, tighter bounds on the metrological precision in the presence of quantum correlations.

The paper is organized as follows. In Sec. \ref{sec:QFI}, we introduce quantum Fisher information and how it is connected to quantum speed of evolution. Sec. \ref{sec:lQFI} is dedicated to local quantum Fisher information and the quantum correlation measure based on it. We revisit the
properties of the quantum correlation measure proposed in \cite{Girolami14}, but from a slightly different approach. The role of quantum correlations in quantum metrology is studied in Sec. \ref{sec:metro} and finally we conclude in Sec. \ref{sec:concl}.

\section{\label{sec:QFI} Quantum Fisher information (QFI)}
Our whole analysis is 
centralized on the QFI which is a Riemannian metric in quantum geometry of state space \cite{Bengtsson06}.
The geometric distance between two arbitrary quantum states, in the projective Hilbert space, using Bures distance 
\cite{Wootters81, Bengtsson06}, is given by $ D_B^2\left( \rho,\sigma \right)= 4\left(1- \sqrt{F_B(\rho,\sigma)}\right)$,
where $F_B(\rho,\sigma)=\left(\mbox{Tr} \sqrt{\sqrt{\rho}\sigma \sqrt{\rho}}\right)^2 $, the Uhlmann fidelity. 
Now consider a smooth dynamical process in the Hilbert space of density matrix
following unitary evolution, parametrized by time $t$. That 
leads to an evolution of the quantum state from $\rho(t_1)$ to $\rho(t_2)$ such that the $D_B\left( \rho(t_1),\rho(t_2)\right)$ 
is a piecewise smooth function of $t_1$ and $t_2$.  
For an infinitesimal time evolution from 
$t$ to $t+dt$ with a given unitary $U=\mbox{exp}[-iHt]$, the distance between the initial and final states, becomes
\begin{equation}
 D_B^2 \left( \rho(t),\rho(t+dt) \right)=dD_B^2=\mathcal{F}^2\left(\rho(t),H \right) dt^2, 
\end{equation}
where we ignore $\mathcal{O}(dt^3)$. The $\mathcal{F}^2\left(\rho(t),H \right)$ is the QFI \cite{Braunstein94}. 
Note that $\mathcal{F}\left(\rho(t),H \right)$ is the instantaneous speed, $\frac{dD_B}{dt}$, of quantum evolution due to the Hamiltonian, $H$, 
in the projective Hilbert space and vanishes iff $[\rho(t),H]=0$. Without loss of generality we assume $\rho(t)\equiv\rho$ in the following.
QFI is defined as $\mathcal{F}^2(\rho, H)=\frac{1}{4}\mbox{Tr}(\rho L^2)$
where $L$ is the symmetric logarithmic derivative (SLD) operator which can be expressed as 
$\frac{d\rho}{dt}=i [\rho, H]=\frac{1}{2} (L\rho + \rho L )$.  QFI, based on SLD operator, has several important 
properties \cite{Braunstein94} inheriting from Bures distance \cite{Wootters81, Bengtsson06}. These properties 
such as convexity, invariance under the unitaries on both initial and final states, and 
monotonicity under completely positive trace preserving (CPTP) maps, establish QFI as the fundamental quantity in
quantum geometry, quantum information and quantum metrology.
For a given quantum state $\rho=\sum_m \lambda_m |m\rangle \langle m |$ with $\sum_m \lambda_m=1$ and $\lambda_m\geqslant0$, 
the QFI reduces to \cite{Braunstein94}
\begin{equation}
 \mathcal{F}^2(\rho,H)=\frac{1}{2}\sum_{m\neq n}\frac{(\lambda_m-\lambda_n)^2}{\lambda_m+\lambda_n}|\langle m | H | n \rangle |^2.
 \label{eq:QFI}
\end{equation}
The summation is carried out 
under the condition $\lambda_m+\lambda_n>0$. 
Henceforth, whenever there appears summation with $\lambda_m+\lambda_n$ in the denominator, we assume that the sum is taken 
under the condition $\lambda_m+\lambda_n>0$ with $\rho=\sum_m \lambda_m |m\rangle \langle m |$, unless otherwise stated.

\begin{figure}
\centering 
\includegraphics[width=0.45\textwidth, angle=0]{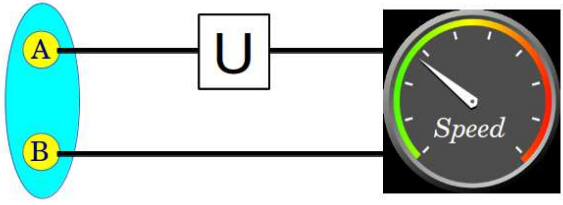}
\caption{\label{fig:SpeedMeter} (Color online). The quantum Fisher information represents the quantum speed of evolution when it is driven with a unitary. The schematics depicts a special case, when one party of a bipartite quantum state, $A$, is driven with a local unitary $U=\mbox{exp}[-iH_At]$. The resultant global quantum speed we call as the local quantum Fisher information. 
}
\end{figure}

\section{\label{sec:lQFI} Local quantum Fisher information and quantum correlation}
%
In this section we review the quantum correlation measure based on QFI originally introduced in \cite{Girolami14}.
Let us consider an $M \times N$ bipartite quantum state $\rho=\sum_m \lambda_m |m\rangle \langle m |$ in the Hilbert space $\mathcal{H}^M_A \otimes \mathcal{H}^N_B$. 
In the case, as shown in Fig. \ref{fig:SpeedMeter}, where the first party, say $A$-party, is driven with the Hamiltonian $H_A=H_a \otimes \mathbb{I}$
the local quantum Fisher Information (lQFI) reduces to: $\mathcal{F}^2(\rho,H_A)= \mbox{Tr} \left(\rho H_A^2 \right) - \sum_{m\neq n}\frac{2 \lambda_m \lambda_n}{\lambda_m+\lambda_n}|\langle m | H_A | n \rangle |^2$ after simple manipulation of Eq. (\ref{eq:QFI}).
Our goal is to quantify quantum correlation in terms of lQFI and for that we define the  
minimum of lQFI, $\mathcal{Q}^2_A(\rho) $, over all local Hamiltonians $\{ H_A \}$ on $A$-party, as
\begin{equation}
\begin{split}
 \mathcal{Q}_{A}^2(\rho) & = \mbox{min}_{\{ H_A \}} \mathcal{F}^2(\rho,H_A).
\end{split}
 \label{eq:wmax}
\end{equation}
The $ \mathcal{Q}_A(\rho)$ represents the minimal quantum speed of evolution, in projective space of 
$\mathcal{H}^M_A \otimes \mathcal{H}^N_B$,  when the party-$A$ is 
driven with the local Hamiltonian $H_A$. Following the properties of QFI \cite{Petz96}, the $ \mathcal{Q}^2_A(\rho)$ acquires several interesting properties: it is nonnegative;
invariant under local unitary operations; 
convex i.e., non increasing under classical mixing and monotonically decreasing 
under local completely positive trace preserving (CPTP) maps on B-party; and in general $\mathcal{Q}_{A}^2(\rho) \neq \mathcal{Q}_{B}^2(\rho)$ except for 
symmetric quantum states. All these important properties indulge us to avow that 
the \textit{ $\mathcal{Q}_{A}^2(\rho)$ is the measure of quantum correlation} and hence, we
introduce the following theorem.


{\bf Theorem I:} \textit{$\mathcal{Q}_{A}^2(\rho)$ vanishes iff the bipartite states
are either classical-quantum (CQ) i.e.,
$\rho=\sum_i p_i |a_i\rangle \langle a_i| \otimes \sigma_{iB}$ where $\langle a_i|a_{i^\prime}\rangle =0$ and 
$\mbox{Tr}(\sigma_{iB}\sigma_{i^\prime B})\neq0$ for $i \neq i^\prime$ 
or classical-classical (CC) i.e.,
$\rho=\sum_i p_i |a_i\rangle \langle a_i| \otimes |b_i\rangle \langle b_i|$ where 
$\langle a_i|a_{i^\prime} \rangle =\langle b_i|b_{i^\prime} \rangle =0$ for $i \neq i^\prime$.}



{\bf Proof:} 
Let us consider a bipartite quantum state $\rho=\sum_{mn} \gamma_{mn} \mathcal{A}_m \otimes \mathcal{B}_n$ represented in 
terms of arbitrary bases of Hermitian operators $\{ \mathcal{A}_m \} \in \mathcal{L}(\mathcal{H}_A^M)$ and 
$\{ \mathcal{B}_n \} \in \mathcal{L}(\mathcal{H}_B^N)$ where $m=1,...,M^2$ and $n=1,...,N^2$. The real valued
$M^2\times N^2$ correlation matrix $[\Gamma]_{mn}=\gamma_{mn}$ can be diagonalized, using singular value decomposition (SVD), to
$U\Gamma V^T=\mbox{diag} [k_1,k_2,...]$ where $U$ and $V$ are the $M^2 \times M^2$ and $N^2 \times N^2$
orthogonal matrices, respectively. Accordingly, the local bases transform to $\mathcal{S}_m=\sum_{m^\prime} U_{mm^\prime} \mathcal{A}_{m^\prime}$ and $\mathcal{R}_n=\sum_{n^\prime} V_{nn^\prime} \mathcal{B}_{n^\prime}$ and the state, in this new basis, 
becomes $\rho=\sum_{m=1}^d k_m \mathcal{S}_m \otimes \mathcal{R}_m$ where $d=\mbox{rank}(\Gamma)$.
For vanishing 
$\mathcal{Q}_{A}^2(\rho)$ the necessary and sufficient condition is $[\rho,H_A]=0$ or equivalently 
$[\mathcal{S}_m,H_a]=0 \ \forall m$. Therefore, the $\{ \mathcal{S}_m \}$ and $H_a$ shares common 
eigenbasis and that is achievable if and only if $[\mathcal{S}_m,\mathcal{S}_n]=0$ for $ m,n=1,...,d$.
So, the states with vanishing $\mathcal{Q}_{A}^2(\rho)$, assumes the form 
$\rho=\sum_{i=1}^M p_i |a_i\rangle \langle a_i| \otimes \sigma_{iB}$ where $\{ |a_i\rangle \}$ are orthonormal
eigenvectors in $\mathcal{H}_A^M$.  $\blacksquare$

Vanishing $\mathcal{Q}_{A}^2(\rho)$ highlights the fact
that there exists at least one local Hamiltonian $H_a$ for which the quantum speed of evolution is zero. On the 
other hand, if the bipartite state is either QC or quantum-quantum (QQ), it is impossible to find a local Hamiltonian $H_a$ with which the quantum states remain stationary in the projective space of $\mathcal{H}_A^M \otimes \mathcal{H}_B^N$ and thus, the $\mathcal{Q}_{A}^2(\rho)$ acquires non-zero value.
Note that, for pure quantum states the $\mathcal{Q}_{A}^2(\rho)$ reduces to the minimal variance over the local unitaries and it 
is an entanglement monotone \cite{Girolami13}.

For a bipartite quantum state of $2\times N$ dimension the $\mathcal{Q}_{A}^2(\rho)$
can be calculated easily. 
We note that the maximally informative Hamiltonians on the $A$-party are the traceless Hermitian 
operators with non-degenerate spectrum. So it is sufficient to consider the Hamiltonian as 
$H_a=\overline{r}.\overline{\sigma}$ where $|r|=1$, i.e.,  
$\overline{r}=\{ \cos \theta, \sin \theta \cos \phi, \sin \theta \sin \phi \}$ and 
$\overline{\sigma}=\{\sigma_x, \sigma_y, \sigma_z  \}$ are the Pauli matrices. 
Now the minimum of lQFI $\mathcal{Q}_{A}^2(\rho)$, which is solely the property of the quantum state, can be 
calculated analytically as
\begin{equation}
\begin{split}
 \mathcal{Q}_{A}^2(\rho) & = \mbox{min}_{\{ \overline{r}\}} \mathcal{F}^2(\rho,H_A), \\
           & = 1- \lambda_w^{max},
 \end{split}
 \label{eq:wmax}
\end{equation}
where $\lambda_w^{max}$ is the largest eigenvalue of the real symmetric matrix 
$\left[W\right]_{ij}= \sum_{m\neq n}\frac{2 \lambda_m \lambda_n}{\lambda_m+\lambda_n} \langle m | \sigma_i \otimes \mathbb{I} | n \rangle \langle n | \sigma_j \otimes \mathbb{I} | m \rangle$.  
Here to minimize $\mathcal{F}^2(\rho,H_A)$, we maximize $\mathbf{r}^{T}W\mathbf{r}$ over all unit vectors $\mathbf{r}$ for the real symmetric matrix $W$ and the maximum value equals to the largest eigenvalue of $\left[W\right]_{ij}$ \cite{Horn90}.
Note that an equivalent formula of $\mathcal{Q}_{A}^2(\rho)$, for $2\times N$ bipartite quantum states, is also derived in \cite{Girolami14}.

\begin{figure}
\centering 
\includegraphics[width=0.45\textwidth, angle=0]{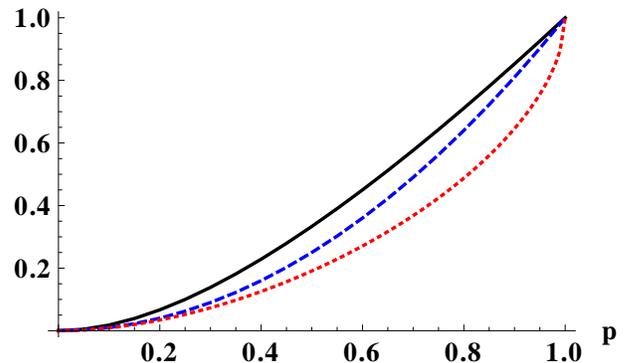}
\caption{\label{fig:QCs} (Color online). The variation quantum correlation measures based on lQFI $\mathcal{Q}_{A}^2(\rho_{AB})$ (black solid trace), Hilbert-Schmidt distance $D^2_{HS}(\rho_{AB})$ (blue dashed trace) and Hellinger distance $D^2_{H}(\rho_{AB})$ (red dotted trace) for two-qubit Werner state $\rho_{AB}=p |\psi^+ \rangle \langle \psi^+| + \frac{1-p}{4}\mathbb{I}_4$ where $|\psi^+ \rangle =\frac{1}{\sqrt{2}}(|00\rangle + |11\rangle)$, with respect to the state parameter $p$.
}
\end{figure}

The $\mathcal{Q}_{A}^2(\rho)$ can be compared with other measures of quantum correlation of a bipartite quantum state $\rho=\sum_m \lambda_m | m \rangle \langle m |$ which are based on various geometric distances such as Hilbert-Schmidt \cite{Dakic10} and Hellinger \cite{Girolami13} distances (see Fig. \ref{fig:QCs}). These measures can again be interpreted as the minimum quantum speed, quantified with the corresponding geometric distances \cite{Girolami13, Giampaolo13}, of local unitary evolutions with the local Hamiltonian $H_A=\overline{r}.\overline{\sigma} \otimes \mathbb{I}$ with $|r|=1$ and $\overline{\sigma}$ are the Pauli matrices. The measure based 
on the Hilbert-Schmidt distance, also known as geometric quantum discord \cite{Giampaolo13, Dakic10}, is given by $D^2_{HS}(\rho)_A=\mbox{min}_{\{ \overline{r}\}} \frac{1}{2}\sum_{m,n}(\lambda_m-\lambda_n)^2|\langle m | H_A | n \rangle |^2$. The local uncertainty measure of quantum correlation, based on Helligner distance \cite{Girolami13}, is written as $D^2_{H} (\rho)_A=\mbox{min}_{\{ \overline{r}\}} \frac{1}{2}\sum_{m,n}(\lambda_m^{1/2}- \lambda_n^{1/2})^2|\langle m | H_A | n \rangle |^2$. From the expression of QFI, we have $\mathcal{Q}_{A}^2(\rho) \geqslant \{D^2_{HS}(\rho)_A,  D^2_{H}(\rho)_A \}$. In Fig. \ref{fig:QCs} we consider an example of a two-qubit Werner state $\rho_{AB}=p |\psi^+ \rangle \langle \psi^+| + \frac{1-p}{4}\mathbb{I}_4$, where $|\psi^+ \rangle =\frac{1}{\sqrt{2}}(|00\rangle + |11\rangle)$ and compare the quantum correlation measures, with respect to the state parameter $p$. The solid (black), dashed (blue) and dotted (red) traces correspond to the quantum correlation measures $\
mathcal{Q}_A^2(\rho_{AB})$, $D^2_{HS}(\rho_{AB})$ and $D^2_{H}(\rho_{AB})$ respectively. The state is entangled for $p>\frac{1}{3}$ while it is quantum correlated for $p>0$. The measure $\mathcal{Q}_A^2(\rho_{AB})$ can reliably quantify it along with other two measures. Though $\mathcal{Q}_A^2(\rho_{AB})$ is appeared to be  a variant of discord-like quantifier of quantum correlation along with $D^2_{HS}(\rho_{AB})$ and $D^2_{H}(\rho_{AB})$, it establishes its precedence over the others in the context of quantum metrology.

\section{\label{sec:metro} Delimiting quantum metrology}
The intimate connection of QFI with quantum metrology and quantum correlation present in a system entices one to ask the question, we pose in this paper, on the role of quantum correlation in quantum metrology. 
In quantum parameter estimation, a quantum state $\rho$, which acts as a probe, undergoes a unitary 
transformation (in general a shift in phase) so that the evolved state becomes 
$\rho_{\theta}=e^{-i \theta H} \rho e^{i \theta H}$, where
$H$ is the Hamiltonian assumed to have non-degenerate spectrum. The parameter $\theta$ is encoded in the 
state $\rho_{\theta}$ and the task is to estimate the unobservable parameter $\theta$. Interestingly, the 
lower bound on the error  (or variance, $\varDelta \theta$), in estimating $\theta$, is independent of the choice 
of the measurements (POVMs) performed after the unitary evolution and solely determined by the dependence 
of the output state on the parameter $\theta$. For a single shot experiment, it is given by the celebrated quantum 
Cram{\'e}r-Rao (qCR) bound \cite{Braunstein94} as $\varDelta \theta \geqslant \frac{1}{\mathcal{F}(\rho, H)} $. 

For simplicity here we consider a bipartite state $\rho$ of $2 \otimes N$ dimension and the parameter estimation
process with local unitary evolution. Though our analysis can easily be extended to the 
systems with arbitrary $M \otimes N$ dimensions. Hence, $\varDelta \theta \geqslant \frac{1}{\mathcal{F}(\rho, H_A)}$,
where $H_A=H_a\otimes\mathbb{I}$ and $H_a$ is the local Hamiltonian acting on the $A$-party. In the presence of quantum correlation, we have 
non-zero $\mathcal{Q}_{A}^2(\rho)$ and lQFI is lower bounded as $\mathcal{Q}_{A}^2(\rho) \leqslant \mathcal{F}(\rho, H_A)$.
The upper bound of lQFI can also be derived and it is the maximal $\mathcal{F}^2(\rho, H_A)$ over 
all possible $H_A$ and can be calculated analytically as
\begin{equation}
\begin{split}
 \mathcal{P}^2_A(\rho) & =\mbox{max}_{\{ \overline{r} \}}  \mathcal{F}^2(\rho, H_A), \\
                       & =1- \lambda_w^{min},
 \end{split}
\end{equation}
where $\lambda_w^{min}$ is the smallest eigenvalue of the real symmetric matrix 
$\left[W\right]_{ij}= \sum_{m,n}\frac{2 \lambda_m \lambda_n}{\lambda_m+\lambda_n} \langle m | \sigma_i \otimes \mathbb{I} | n \rangle \langle n | \sigma_j \otimes \mathbb{I} | m \rangle$.
The maximization of $\mathcal{F}^2(\rho, H_A)$ is carried out using the same logic as in Eq. (\ref{eq:wmax}) except that here we consider the smallest eigenvalue of $W$.
The $\mathcal{P}^2_A(\rho)$ possesses all the good properties as of $\mathcal{Q}^2_A(\rho)$. Now the bounds on 
the lQFI becomes $\mathcal{F}^2_m(\rho) \geqslant \mathcal{F}^2(\rho,H_A) \geqslant \mathcal{Q}_A^2(\rho)$. 
Hence in the presence of quantum correlation  $(\mathcal{Q}_A(\rho)\neq0)$, the error on the estimated 
parameter, in a single shot experiment, is given by
\begin{equation}
 \begin{split}
\frac{1}{\mathcal{Q}_A(\rho)}  \geqslant \varDelta \theta \geqslant \frac{1}{\mathcal{P}_A(\rho)}.
\end{split}
\end{equation}
Remarkably the quantum correlated states have intrinsic precision in metrology 
with local unitaries that is inverse to the quantum correlation present in the system, while it is absent for the CQ states. This intrinsic precision is also tested experimentally \cite{Girolami14}. On the other hand the maximum metrological precision achievable, given all choices of local unitaries, is determined by the inverse of $\mathcal{P}_A(\rho)$.


\begin{figure}
\centering 
\includegraphics[width=0.35\textwidth, angle=0]{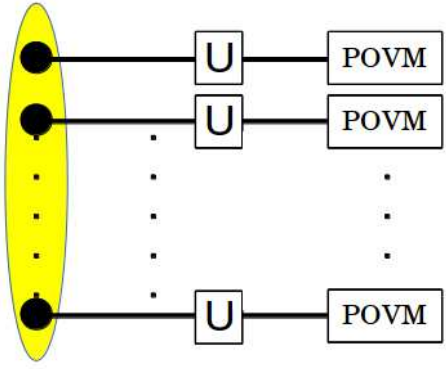}
\caption{\label{fig:Metrology} (Color online). A schematic of a quantum metrological experiment with N-party quantum state $\rho_N$, which acts as a probe. The $\rho_N$ is driven with local unitaries and as a result it encodes the unobservable parameter, say $\theta$ (in general a shift in phase), which is to be estimated with measurements (POVMs). 
}
\end{figure}

It is known that the QFI is additive quantity on product states and in particular $\mathcal{F}^2(\rho^{\otimes N}, \oplus _1^N H)= N \mathcal{F}^2(\rho^{\otimes N}, H)$. Thus, the qCR bound becomes $\varDelta \theta \geqslant \frac{1}{ \sqrt{N}\mathcal{F}(\rho^{\otimes N}, H)} $. The term $\sqrt{N}$ in the denominator may be equivalently interpreted as due to $N$ independent repetitions of an experiment with a state $\rho$ or a single shot experiment (see Fig. \ref{fig:Metrology}) with a multi-party state $\rho_N=\rho^{\otimes N}$. The corresponding scaling of metrological error $\varDelta \theta \sim \mathcal{O}(\frac{1}{\sqrt{N}})$ is referred to as the SQL. The situation becomes very different when the system $\rho_N$ is quantum correlated. The $N$-party state $\rho_N$ is referred to as the \textit{genuinely quantum correlated} (GQC) state if $\mathcal{Q}_k(\rho_N)\neq 0 \ \forall k \in \{1...N \}$. Now we show that for a GQC state, $\rho_N$, the $\mathcal{F}^2(\rho_N, \oplus _{i=1}^N H_i) \neq \sum_i \
\mathcal{F}^2(\rho_N, H_i)$.
Consider an $M\times N$ bipartite system $\rho$ is driven with the Hamiltonian $H=H_A+H_B$.
In such case, the global (or joint) quantum Fisher information (gQFI) becomes 
\begin{equation}
\begin{split}
 \mathcal{F}^2(\rho,H_A+H_B)= & \mathcal{F}^2(\rho,H_A)+\mathcal{F}^2(\rho,H_B) \\
                            & + 2 \mathcal{C}(\rho, H_A, H_B),
 \end{split}
 \label{eq:GQFI}
\end{equation}
where the third term (also called the interference term)
$\mathcal{C}(\rho, H_A, H_B)= \frac{1}{2}\sum_{m \neq n}\frac{(\lambda_m-\lambda_n)^2}{\lambda_m+\lambda_n} \langle m | H_A 
| n \rangle    \langle n | H_B | m \rangle$. 

{\bf Theorem II:} \textit{For a non-GQC states, the $\mathcal{C}(\rho, H_A, H_B)$ vanishes for 
arbitrary $H_A$ and $H_B$.}

{\bf Proof:} 
To prove the above theorem it is sufficient to show that every non-diagonal ($m \neq n$) elements $C_{mn}=\langle m | H_A | n \rangle \langle n | H_B | m \rangle$, of $\mathcal{C}(\rho, H_A, H_B)$ in Eq. (\ref{eq:GQFI}), is vanishing as long as it is a non-GQC state. Let us consider an $M \times N$ bipartite non-GQC state, say CQ state $\rho =\sum_{ij} q_{ij} |a_{i}  b_{ij}\rangle \langle a_{i}  b_{ij} | $ with $|a_{i}  b_{ij}\rangle = |a_{i}\rangle \otimes | b_{ij}\rangle$. Here $\{| m \rangle, | n \rangle \}=\{| a_{i} b_{ij} \rangle, | a_{i^\prime} b_{i^\prime j^\prime} \rangle\}$, where $\langle a_{i}| a_{i^\prime}\rangle=0$  $\forall i\neq i^\prime$ and $\langle b_{ij}| b_{i^\prime j^\prime}\rangle=0$ for $ i= i^\prime$ and $ j\neq j^\prime$ only. Now in the new basis the non-diagonal terms become $C_{mn}=\langle a_{i}|H_a|a_{i^\prime}\rangle     \langle b_{ij} |b_{i^\prime j^\prime}\rangle  \langle a_{i^\prime} | a_{i} \rangle \langle b_{i^\prime j^\prime}|H_b|b_{ij}\rangle$. To assure $m\neq n$, we are 
left 
with three choices. First, $i= i^\prime$ but $j\neq j^\prime$. Second, $i\neq i^\prime$ but $j = j^\prime$ and finally, $i\neq i^\prime$ but $j \neq j^\prime$. Interestingly, in all three cases the $C_{mn}=0$ and thus the $\mathcal{C}(\rho, H_A, H_B)=0$, for arbitrary $H_A$ and $H_B$. Similarly, it is straightforward to show that the $\mathcal{C}(\rho, H_A, H_B)$ also vanishes for QC and CC states. $\blacksquare$

\begin{figure}
\centering 
\includegraphics[width=0.47\textwidth, angle=0]{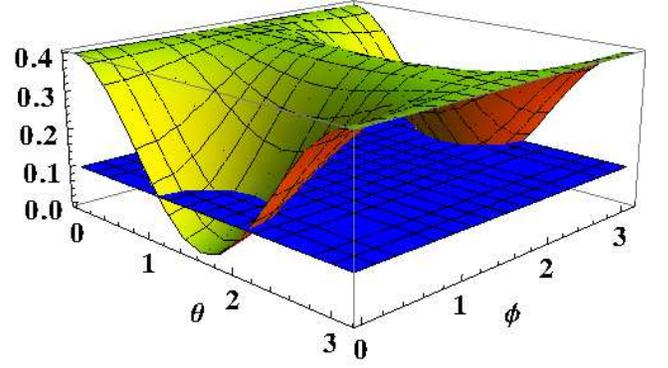}
\caption{\label{fig:Interfare} (Color online). The figure displays the role of \textit{genuine quantum correlation}, other than entanglement, in increasing quantum Fisher information. The figure is calculated for the bipartite quantum state, the Werner state $\rho=p |\psi^+ \rangle \langle \psi^+| + \frac{1-p}{4}\mathbb{I}_4$ where $|\psi^+ \rangle =\frac{1}{\sqrt{2}}(|00\rangle + |11\rangle$) and  $p=\frac{1}{4}$, which has zero entanglement. The parameters $\theta$ and $\phi$ belong to the local Hamiltonians $H_a=H_b=\overline{r}.\overline{\sigma}$, where $\overline{r}=\{ \cos \theta, \sin \theta \cos \phi, \sin \theta \sin \phi \}$ and $\overline{\sigma}=\{\sigma_x, \sigma_y, \sigma_z  \}$ are the Pauli matrices. The vertical axis (the z-axis) represents the quantum Fisher information. the flat  surface (in blue) represents lQFI (which is independent of $\theta$ and $\phi$) $\mathcal{F}(\rho,H_A)=0.1$. The curved surface (in yellow) corresponds to the gQFI. For $\theta=\frac{\pi}{2}, \ \phi=0$, gQFI 
vanishes due to destructive interference while for $\theta=0, \ \phi=0$, it acquires maximum value $4\mathcal{F}(\rho,H_A)=0.4$ due to constructive interference between the local unitary evolutions.}
\end{figure}


It is clear from Theorem II that for non-GQC states, the gQFI is, just, simple algebraical sum of the lQFIs, as the interference term becomes zero. While for GQC states the interference term 
$\mathcal{C}(\rho, H_A, H_B)$, is non-vanishing and can acquire both positive and negative values. 
So it reveals that genuine quantum correlation plays a pivotal role to increase as well as to reduce gQFI than that of the sum of lQFIs (e.g., see Fig. \ref{fig:Interfare}). Therefore for the GQC states, we have 
$-\mathcal{F}(\rho,H_A)\mathcal{F}(\rho,H_B) \leqslant \mathcal{C}(\rho, H_A, H_B) \leqslant \mathcal{F}(\rho,H_A)\mathcal{F}(\rho,H_B)$. As a result the gQFI is upper and lower bounded as $\left( \mathcal{F}(\rho,H_A)-  \mathcal{F}(\rho,H_B)  \right)^2  \leqslant \mathcal{F}^2  (\rho,H_A+H_B) \leqslant \left( \mathcal{F}(\rho,H_A)+\mathcal{F}(\rho,H_B) \right)^2 $, where the equalities can be achieved for certain choices of $H_A$ and $H_B$ depending on the state $\rho$. This particular feature can be adjusted to attain better metrological precision beyond SQL, even to reach HL.
In Fig. \ref{fig:Interfare}, we consider an example of Werner state with zero entanglement, $\rho=p |\psi^+ \rangle \langle \psi^+| + \frac{1-p}{4}\mathbb{I}_4$, where $|\psi^+ \rangle =\frac{1}{\sqrt{2}}(|00\rangle + |11\rangle$) and $p=\frac{1}{4}$. It is a symmetric state and $\mathcal{P}^2_i(\rho)=\mathcal{P}^2(\rho)$ for $i=A,B$. For the Hamiltonian $H_a=H_b=\overline{r}.\overline{\sigma}$, where $\overline{r}=\{ \cos \theta, \sin \theta \cos \phi, \sin \theta \sin \phi \}$ and $\overline{\sigma}=\{\sigma_x, \sigma_y, \sigma_z  \}$ the Pauli matrices,  gQFI vanishes for $\theta=\frac{\pi}{2}, \ \phi=0$. The lQFI is independent of $\theta$ and $\phi$, i.e., $ \mathcal{F}^2(\rho,H_A)= \mathcal{P}^2(\rho)=\mathcal{Q}^2(\rho)=0.1$. While for $\theta=0, \ \phi=0$, the gQFI becomes maximum and $\mathcal{F}^2(\rho,H_A+H_B)=4\mathcal{P}^2(\rho)=0.4$. 

For arbitrary bipartite states and Hamiltonians, the tighter bounds on gQFI can be provided since $\left( \mathcal{F}(\rho,H_A)-  \mathcal{F}(\rho,H_B)  \right)^2 \geqslant \left( \mathcal{Q}_A(\rho)-  \mathcal{Q}_B(\rho)  \right)^2$ and $\left( \mathcal{F}(\rho,H_A)+\mathcal{F}(\rho,H_B)  \right)^2 \leqslant \left( \mathcal{P}_A(\rho)+\mathcal{P}_B(\rho)  \right)^2$. As a result the bounds on the error in parameter estimation, with GQC states, is 
 \begin{equation}
\begin{split}
\frac{1}{\mathcal{Q}_A(\rho)-  \mathcal{Q}_B(\rho)} \geqslant  \varDelta \theta 
                                    \geqslant \frac{1}{\mathcal{P}_A(\rho) + \mathcal{P}_B(\rho)} 
\end{split}
\end{equation}
For N-party symmetric  GQC states and identical local Hamiltonians, 
 i.e., $H_i=\otimes_1^{i-1} \mathbb{I}\otimes H \otimes_{i+1}^{N} \mathbb{I}$, the  $\mathcal{F}(\rho_N,H_i)=\mathcal{F}(\rho_N,H)$ $\forall i$. Similarly, 
 $\mathcal{Q}_i(\rho)=\mathcal{Q}(\rho)$ and $\mathcal{P}_i(\rho)=\mathcal{P}(\rho)$  $\forall i$. 
 It is straightforward to see that for N-party symmetric GQC states $\rho_N$, gQFI is 
$\mathcal{F}^2(\rho_N, \oplus_{i=1}^N H_i)=N^2 \mathcal{F}^2(\rho_N,H)$ for the particular choice of quantum states and the local
Hamiltonians, which is not the case otherwise. Note, here we need quantum correlation, which may not include entanglement, to attain HL with the scaling $\varDelta \theta \sim \mathcal{O}\left(\frac{1}{N} \right)$. Further, the tighter qCR bound, independent of the local unitaries, can be given  in terms of solely a state dependent quantity as 
  \begin{equation}
   \varDelta \theta \geqslant \frac{1}{N\mathcal{P}_A(\rho)}.
  \end{equation}
It is interesting to note that for a class of GQC states with arbitrary $H_i=\otimes_1^{i-1} \mathbb{I}\otimes H \otimes_{i+1}^{N} \mathbb{I}$, the interference term, in Eq. (\ref{eq:GQFI}), reduces to $\mathcal{C}(\rho, H_i, H_j)=-\mathcal{F}^2(\rho,H)$ and consequently gQFI vanishes arbitrarily, e.g., the Werner state of the form $\rho=\frac{1-p}{4}\mathbb{I}_4 +|\phi^- \rangle \langle \phi^- |$ where $|\phi^- \rangle =\frac{1}{\sqrt{2}}\left(|01\rangle - |10\rangle \right)$. Hence, these quantum states are useless from the perspectives of quantum metrology.


\section{\label{sec:concl} Conclusion}
In this paper we investigate the role of quantum correlation in quantum metrology. The best suited measure of quantum correlation, which laid the basis of investigation, is the one based on quantum Fisher information originally introduced in \cite{Girolami14}. As the quantum Fisher information represents the quantum speed of evolution, when the geometric distance in the quantum state space is expressed in terms of Bures metric, the quantum correlation can be interpreted from quantum dynamical perspective. Thus the measure of quantum correlation becomes identified with the minimal speed of evolution over all possible local unitaries.
This dynamical approach to quantify quantum correlation provides us the premise to study quantum metrology in the presence of quantum correlation.  The quantum correlation measure gives the lowest bound on the error in quantum metrology, when the experiment is performed on one of the sub-systems. An upper bound on the error is also derived analytically. For quantum metrology with  multi-party quantum systems, we show that not only the entanglement but also the quantum correlation plays important roles in enhancing precision in quantum metrology. Even in the absence of entanglement, quantum correlation can be exploited to go beyond SQL and attain HL. The very reason for better precision is the constructive interference between the local unitary evolutions, resulting in 
larger quantum Fisher information and that happens only in the presence of quantum correlation. It is also 
possible to have destructive interference, in the presence of quantum correlation, to make the resultant quantum Fisher information very small or even zero. We have derived the tighter bounds on the metrological error in the presence of quantum correlation in multiparty scenario. 

{\bf Acknowledgement} -- The author thanks Prof. Arun K. Pati and Himadri S. Dhar for carefully reading the manuscript. The author also gratefully acknowledges the useful comments by Gerardo Adesso and Davide Girolami.

\end{document}